\documentclass[12pt,a4paper]{article}
\makeatletter
\def\eqnarray{%
\stepcounter{equation}%
\let\@currentlabel=\theequation
\global\@eqnswtrue
\global\@eqcnt\z@
\tabskip\@centering
\let\\=\@eqncr
$$\halign to \displaywidth\bgroup\@eqnsel\hskip\@centering
$\displaystyle\tabskip\z@{##}$&\global\@eqcnt\@ne
\hfil$\displaystyle{{}##{}}$\hfil
&\global\@eqcnt\tw@$\displaystyle\tabskip\z@{##}$\hfil
\tabskip\@centering&\llap{##}\tabskip\z@\cr}
\makeatother
\newcommand{\ket}[1]{{\vert{#1}\rangle}}
\newcommand{\bra}[1]{{\langle{#1}\vert}}
\newcommand{\kett}[2]{{\vert{#1,#2}\rangle}}
\newcommand{\braa}[2]{{\langle{#1,#2}\vert}}
\newcommand{\calh}{{\cal H}}

\newcommand{\fukuso}{{\mathbf C}}

\newcommand{\futon}{{\bf N}}

\newcommand{\zettai}[1]{{\vert{#1}\vert}}

\begin{document}

\title{\sl Coherent States and Some Topics in Quantum 
Information Theory : Review}
\author{
  Kazuyuki FUJII
  \thanks{E-mail address : fujii@yokohama-cu.ac.jp }\\
  Department of Mathematical Sciences\\
  Yokohama City University\\
  Yokohama, 236-0027\\
  Japan
  }
\date{}
\maketitle
%
%
%
%
\begin{abstract}
  In the first half we make a short review of coherent states and 
  generalized coherent ones based on Lie algebras su(2) and su(1,1), 
  and the Schwinger's boson method to construct representations of the 
  Lie algebras. 
  In the second half we make a review of recent developments on 
  both swap of coherent states and cloning of coherent states which are 
  important subjects in Quantum Information Theory. 
\end{abstract}
%


%
%
%
%

\section{Introduction}

The purpose of this paper is to introduce several basic theorems of 
coherent states and generalized coherent states based on Lie algebras 
su(2) and su(1,1), and to give some applications of them to Quantum 
Information Theory. 

In the first half we make a general review of coherent states and 
generalized coherent states based on Lie algebras su(2) and su(1,1). 

Coherent states or generalized coherent states play an important role in 
quantum physics, in particular, quantum optics, see \cite{KS} and its 
references, or the books \cite{MW}, \cite{WPS}. 
They also play an important one in mathematical physics, 
see the textbook \cite{AP}. For example, they are very useful in performing 
stationary phase approximations to path integral, \cite{FKSF1}, 
\cite{FKSF2}, \cite{FKS}. 

In the latter half we apply a method of generalized coherent states 
to some important topics in Quantum Information Theory, in particular, 
swap of coherent states and cloning of coherent ones. 

Quantum Information Theory is one of most exciting fields in modern physics 
or mathematical physics or applied mathematics. 
It is mainly composed of three subjects 
\begin{center}
Quantum Computation,\quad Quantum Cryptgraphy\quad and\quad 
Quantum Teleportation.
\end{center}
See for example \cite{LPS}, \cite{ASt} or \cite{KF1}, \cite{KF7}, \cite{KF8}. 
Coherent states or generalized coherent states also play an important role 
in it. 

We construct the swap operator of coherent states by making use of 
a generalized coherent operator based on su(2) and moreover show 
an ``imperfect cloning" of coherent states, and last present 
some related problems.

\section{Coherent and Generalized Coherent Operators Revisited}

We make a some review of general theory of both a coherent operator and 
generalized coherent ones based on Lie algebras $su(1,1)$ and $su(2)$.

\subsection{Coherent Operator}
Let $a(a^\dagger)$ be the annihilation (creation) operator of the harmonic 
oscillator.
If we set $N\equiv a^\dagger a$ (:\ number operator), then
\begin{equation}
  \label{eq:2-1-1}
  [N,a^\dagger]=a^\dagger\ ,\
  [N,a]=-a\ ,\
  [a^\dagger, a]=-\mathbf{1}\ .
\end{equation}
Let $\calh$ be a Fock space generated by $a$ and $a^\dagger$, and
$\{\ket{n}\vert\  n\in\futon\cup\{0\}\}$ be its basis.
The actions of $a$ and $a^\dagger$ on $\calh$ are given by
\begin{equation}
  \label{eq:2-1-2}
  a\ket{n} = \sqrt{n}\ket{n-1}\ ,\
  a^{\dagger}\ket{n} = \sqrt{n+1}\ket{n+1}\ ,
  N\ket{n} = n\ket{n}
\end{equation}
where $\ket{0}$ is a normalized vacuum ($a\ket{0}=0\  {\rm and}\  
\langle{0}\vert{0}\rangle = 1$). From (\ref{eq:2-1-2})
state $\ket{n}$ for $n \geq 1$ are given by
\begin{equation}
  \label{eq:2-1-3}
  \ket{n} = \frac{(a^{\dagger})^{n}}{\sqrt{n!}}\ket{0}\ .
\end{equation}
These states satisfy the orthogonality and completeness conditions
\begin{equation}
  \label{eq:2-1-4}
   \langle{m}\vert{n}\rangle = \delta_{mn}\quad \mbox{and}\quad 
   \sum_{n=0}^{\infty}\ket{n}\bra{n} = \mathbf{1}. 
\end{equation}
\noindent{\bfseries Definition}\quad We call a state defined by 
\begin{equation}
\label{eq:2-1-7}
\ket{z} =  \mbox{e}^{za^{\dagger}- \bar{z}a}\ket{0}\equiv D(z)\ket{0} 
\quad \mbox{for}\quad z\ \in\ \fukuso 
\end{equation}
the coherent state.

\subsection{Generalized Coherent Operator Based on $su(1,1)$}
We consider a spin $K\ (> 0)$ representation of $su(1,1) 
\subset sl(2,\fukuso)$ and set its generators 
$\{ K_{+}, K_{-}, K_{3} \}$ $((K_{+})^{\dagger} = K_{-})$, 
\begin{equation}
  \label{eq:2-2-3}
 [K_{3}, K_{+}]=K_{+}, \quad [K_{3}, K_{-}]=-K_{-}, 
 \quad [K_{+}, K_{-}]=-2K_{3}.
\end{equation}
We note that this (unitary) representation is necessarily infinite 
dimensional. 
The Fock space on which $\{ K_{+}, K_{-}, K_{3} \}$ act is 
$\calh_K \equiv \{\kett{K}{n} \vert n\in\futon\cup\{0\} \}$ and 
whose actions are
\begin{eqnarray}
  \label{eq:2-2-4}
 K_{+} \kett{K}{n} &=& \sqrt{(n+1)(2K+n)}\kett{K}{n+1},\quad 
 K_{-} \kett{K}{n}  =  \sqrt{n(2K+n-1)}\kett{K}{n-1},  \nonumber  \\
 K_{3} \kett{K}{n} &=& (K+n)\kett{K}{n}, 
\end{eqnarray}
where $\kett{K}{0}$ is a normalized vacuum ($K_{-}\kett{K}{0}=0$ and 
$\langle K,0|K,0 \rangle =1$). We have written $\kett{K}{0}$ instead 
of $\ket{0}$  to emphasize the spin $K$ representation, see \cite{FKSF1}. 
From (\ref{eq:2-2-4}), states $\kett{K}{n}$ are given by 
\begin{equation}
  \label{eq:2-2-5}
 \kett{K}{n} =\frac{(K_{+})^n}{\sqrt{n!(2K)_n}}\kett{K}{0} ,
\end{equation}
where $(a)_n$ is the Pochammer's notation
$
(a)_n \equiv  a(a+1) \cdots (a+n-1).
$
These states satisfy the orthogonality and completeness conditions 
\begin{equation}
  \label{eq:2-2-7}
  \langle K,m \vert K,n \rangle =\delta_{mn}\quad \mbox{and}\quad 
  \sum_{n=0}^{\infty}\kett{K}{n}\braa{K}{n}\ = \mathbf{1}_K.
\end{equation}
Now let us consider a generalized version of coherent states : 

\noindent{\bfseries Definition}\quad We call a state defined by 
\begin{equation}
   \label{eq:2-2-8}
 \ket{z} = \mbox{e}^{zK_{+} - \bar{z}K_{-}}\kett{K}{0} 
  \quad \mbox{for} \quad z \in \fukuso.
\end{equation}
the generalized coherent state (or the coherent state of Perelomov's 
type based on $su(1,1)$ in our terminology).

Here let us construct an example of this representation. 
We set
\begin{equation}
  \label{eq:2-2-12}
  K_{+}\equiv{1\over2}\left( a^{\dagger}\right)^2\ ,\
  K_{-}\equiv{1\over2}a^2\ ,\
  K_{3}\equiv{1\over2}\left( a^{\dagger}a+{1\over2}\right)\ ,
\end{equation}
then it is easy to check the relations (\ref{eq:2-2-3}).\  
That is, the set $\{K_{+},K_{-},K_{3}\}$ gives a unitary representation of 
$su(1,1)$ with spin $K = 1/4\ \mbox{and}\ 3/4$. We also call an operator 
\begin{equation}
  \label{eq:2-2-14}
   S(z) = \mbox{e}^{\frac{1}{2}\{z(a^{\dagger})^2 - \bar{z}a^2\}}
   \quad \mbox{for} \quad z \in \fukuso 
\end{equation}
the squeezed operator, see the textbook \cite{AP}.

\subsection{Generalized Coherent Operator Based on $su(2)$}
We consider a spin $J\ (> 0)$ representation of $su(2) 
\subset sl(2,\fukuso)$ and set its generators 
$\{ J_{+}, J_{-}, J_{3} \}\ ((J_{+})^{\dagger} = J_{-})$, 
\begin{equation}
  \label{eq:2-3-3}
 [J_{3}, J_{+}]=J_{+}, \quad [J_{3}, J_{-}]=-J_{-}, 
 \quad [J_{+}, J_{-}]=2J_{3}.
\end{equation}
We note that this (unitary) representation is necessarily finite 
dimensional. 
The Fock space on which $\{ J_{+}, J_{-}, J_{3} \}$ act is 
$\calh_{J} \equiv \{\kett{J}{n} \vert 0 \le n \le 2J \}$ and 
whose actions are
\begin{eqnarray}
  \label{eq:2-3-4}
 J_{+} \kett{J}{n} &=& \sqrt{(n+1)(2J-n)}\kett{J}{n+1},\quad 
 J_{-} \kett{J}{n}  =  \sqrt{n(2J-n+1)}\kett{J}{n-1},  \nonumber  \\
 J_{3} \kett{J}{n} &=& (-J+n)\kett{J}{n}, 
\end{eqnarray}
where $\kett{J}{0}$ is a normalized vacuum ($J_{-}\kett{J}{0}=0$ and 
$\langle J,0|J,0 \rangle =1$). We have written $\kett{J}{0}$ instead 
of $\ket{0}$ to emphasize the spin $J$ representation, see \cite{FKSF1}. 
From (\ref{eq:2-3-4}), states $\kett{J}{n}$ are given by 
\begin{equation}
  \label{eq:2-3-5}
 \kett{J}{n} =\frac{(J_{+})^n}{\sqrt{n!{}_{2J}P_n}}\kett{J}{0}.
\end{equation}
These states satisfy the orthogonality and completeness conditions 
\begin{equation}
  \label{eq:2-3-6}
  \langle J,m \vert J,n \rangle =\delta_{mn}\quad \mbox{and}\quad 
  \sum_{n=0}^{2J}\kett{J}{n}\braa{J}{n}\ = \mathbf{1}_{J}.
\end{equation}
Now let us consider a generalized version of coherent states : 

\noindent{\bfseries Definition}\quad We call a state defined by 
\begin{equation}
   \label{eq:2-3-7}
 \ket{z} = \mbox{e}^{zJ_{+} - \bar{z}J_{-}}\kett{J}{0} 
  \quad \mbox{for} \quad z \in \fukuso.
\end{equation}
the generalized coherent state (or the coherent state of Perelomov's 
type based on $su(2)$ in our terminology).

\subsection{Schwinger's Boson Method}
Here let us construct the spin $K$ and $J$ representations by making 
use of Schwinger's boson method.

Next we consider the system of two-harmonic oscillators. If we set
\begin{equation}
  \label{eq:2-4-1}
  a_1 = a \otimes 1,\  {a_1}^{\dagger} = a^{\dagger} \otimes 1;\ 
  a_2 = 1 \otimes a,\  {a_2}^{\dagger} = 1 \otimes a^{\dagger},
\end{equation}
then it is easy to see 
\begin{equation}
  \label{eq:2-4-2}
 [a_i, a_j] = [{a_i}^{\dagger}, {a_j}^{\dagger}] = 0,\ 
 [a_i, {a_j}^{\dagger}] = \delta_{ij}, \quad i, j = 1, 2. 
\end{equation}
We also denote by $N_{i} = {a_i}^{\dagger}a_i$ number operators.

Now we can construct representation of Lie algebras $su(2)$ and $su(1,1)$ 
making use of Schwinger's boson method, see \cite{FKSF1}, \cite{FKSF2}. 
Namely if we set 
\begin{eqnarray}
  \label{eq:2-4-3-1}
  su(2) &:&\quad
     J_+ = {a_1}^{\dagger}a_2,\ J_- = {a_2}^{\dagger}a_1,\ 
     J_3 = {1\over2}\left({a_1}^{\dagger}a_1 - {a_2}^{\dagger}a_2\right), \\
  \label{eq:2-4-3-2}
  su(1,1) &:&\quad
     K_+ = {a_1}^{\dagger}{a_2}^{\dagger},\ K_- = a_2 a_1,\ 
     K_3 = {1\over2}\left({a_1}^{\dagger}a_1 + {a_2}^{\dagger}a_2  + 1\right),
\end{eqnarray}
then it is easy to check that (\ref{eq:2-4-3-1}) satisfies the relations 
(\ref{eq:2-2-3}), while (\ref{eq:2-4-3-2}) satisfies (\ref{eq:2-2-4}). 

In the following we define (unitary) generalized coherent operators 
based on Lie algebras $su(2)$ and $su(1,1)$. 

\noindent{\bfseries Definition}\quad We set 
\begin{eqnarray}
  \label{eq:2-4-5-1}
  su(2) &:&\quad 
U_{J}(z) = e^{z{a_1}^{\dagger}a_2 - \bar{z}{a_2}^{\dagger}a_1}\quad 
  {\rm for}\  z \in \fukuso , \\
  \label{eq:2-4-5-2}
  su(1,1) &:&\quad 
U_{K}(z) = e^{z{a_1}^{\dagger}{a_2}^{\dagger} - \bar{z}a_2 a_1}\quad 
  {\rm for}\  z \in \fukuso.
\end{eqnarray}
For the details of $U_{J}(z)$ and $U_{K}(z)$ see \cite{AP} and \cite{FKSF1}.

\par \vspace{3mm} 
Before closing this section let us make some mathematical preliminaries 
for the latter sections. We have easily
\begin{eqnarray}
  \label{eq:J-rotation-1}
  U_{J}(t)a_{1}U_{J}(t)^{-1}&=&
      cos(\zettai{t})a_{1}-\frac{tsin(\zettai{t})}{\zettai{t}}a_{2}, \\
  \label{eq:J-rotation-2}
  U_{J}(t)a_{2}U_{J}(t)^{-1}&=&
      cos(\zettai{t})a_{2}+\frac{{\bar t}sin(\zettai{t})}{\zettai{t}}a_{1},
\end{eqnarray}
so the map 
$
(a_{1},a_{2}) \longrightarrow 
          (U_{J}(t)a_{1}U_{J}(t)^{-1},U_{J}(t)a_{2}U_{J}(t)^{-1}) 
$
is 
\[
(U_{J}(t)a_{1}U_{J}(t)^{-1},U_{J}(t)a_{2}U_{J}(t)^{-1})=
(a_{1},a_{2})
\left(
  \begin{array}{cc}
     cos(\zettai{t})& \frac{{\bar t}sin(\zettai{t})}{\zettai{t}}\\
     -\frac{tsin(\zettai{t})}{\zettai{t}}& cos(\zettai{t})
   \end{array}
 \right).
\]
We note that 
\[
\left(
  \begin{array}{cc}
     cos(\zettai{t})& \frac{{\bar t}sin(\zettai{t})}{\zettai{t}}\\
     -\frac{tsin(\zettai{t})}{\zettai{t}}& cos(\zettai{t})
   \end{array}
 \right) \in SU(2).
\]

\par \noindent 
On the other hand we have easily
\begin{eqnarray}
  \label{eq:K-rotation-1}
  U_{K}(t)a_{1}U_{K}(t)^{-1}&=&
  cosh(\zettai{t})a_{1}-\frac{tsinh(\zettai{t})}{\zettai{t}}{a_2}^{\dagger}, \\
  \label{eq:K-rotation-2}
  U_{K}(t){a_2}^{\dagger}U_{K}(t)^{-1}&=&
  cosh(\zettai{t}){a_2}^{\dagger}-\frac{{\bar t}sinh(\zettai{t})}{\zettai{t}}
  a_{1},
\end{eqnarray}
so the map 
$
(a_{1},{a_2}^{\dagger}) \longrightarrow 
   (U_{K}(t)a_{1}U_{K}(t)^{-1},U_{K}(t){a_2}^{\dagger}U_{K}(t)^{-1}) 
$
is 
\[
(U_{K}(t)a_{1}U_{K}(t)^{-1},U_{K}(t){a_2}^{\dagger}U_{K}(t)^{-1})=
(a_{1},{a_2}^{\dagger})
\left(
  \begin{array}{cc}
     cosh(\zettai{t})& -\frac{{\bar t}sinh(\zettai{t})}{\zettai{t}}\\
     -\frac{tsinh(\zettai{t})}{\zettai{t}}& cosh(\zettai{t})
   \end{array}
 \right).
\]
We note that 
\[
\left(
  \begin{array}{cc}
     cosh(\zettai{t})& -\frac{{\bar t}sinh(\zettai{t})}{\zettai{t}}\\
     -\frac{tsinh(\zettai{t})}{\zettai{t}}& cosh(\zettai{t})
   \end{array}
 \right) \in SU(1,1). 
\]

\section{Some Topics in Quantum Information Theory}

In this section we don't introduce a general theory of quantum information 
theory (see for example \cite{LPS}), but focus our mind on special topics 
in it, that is, 
\begin{itemize}
   \item swap of coherent states 
   \item cloning of coherent states 
\end{itemize}
Because this is just a good one as examples 
of applications of coherent and generalized coherent states, and 
our method developed in the following may open a new possibility in 
quantum information theory. 

\par \noindent 
First let us define a swap operator : 
\begin{equation}
S : {\calh}\otimes {\calh} \longrightarrow {\calh}\otimes {\calh}, \quad 
  S(a\otimes b)=b\otimes a \quad \mbox{for any}\ a, b \in {\calh} 
\end{equation}
where ${\calh}$ is the Fock space in Section 2.

\par \noindent 
It is not difficult to construct this operator in a universal manner, 
see \cite{KF7} ; Appendix C. 
But for coherent states we can construct a better one by making use of 
generalized coherent operators in the preceding section. 

\par \vspace{3mm} \noindent 
Next let us introduce no cloning theorem, \cite{WZ}.  For that 
we define a cloning (copying) operator C which is unitary 
\begin{equation}
 C : {\calh}\otimes {\calh} \longrightarrow {\calh}\otimes {\calh}, \quad 
    C(h\otimes \ket{0})=h\otimes h \quad \mbox{for any}\ h \in {\calh}\ .
\end{equation}
It is very known that there is no cloning theorem 

\noindent{\bfseries ``No Cloning Theorem''}\quad We have no $C$ above. 

The proof is very easy (almost trivial). Because $2h=h+h \in {\calh}$ and $C$ 
is a linear operator, so 
\begin{equation}
 \label{eq:C-equality}
   C(2h\otimes \ket{0})=2C(h\otimes \ket{0}). 
\end{equation}
The LHS of (\ref{eq:C-equality}) is 
\ \ 
$
   C(2h\otimes \ket{0})= 2h\otimes 2h= 4(h\otimes h),
$
\ \ 
while the RHS of (\ref{eq:C-equality}) is 
\ \ 
$
   2C(h\otimes \ket{0})= 2(h\otimes h).
$
\ \ 
This is a contradiction.\ This is called no cloning theorem. 

\par \noindent 
Let us return to the case of coherent states. 
For coherent states $\ket{\alpha}$ and $\ket{\beta}$\ \ the superposition 
$\ket{\alpha}+\ket{\beta}$ is no longer a coherent state, so that coherent 
states may not suffer from the theorem above. 

\begin{flushleft}
 {\bf Problem}\ \ Is it possible to clone coherent states ? 
\end{flushleft}

\par \noindent
At this stage it is not easy, so we will make do with 
approximating it (imperfect cloning in our terminology) 
instead of making a perfect cloning. 

\par \noindent
We write notations once more. 
\begin{eqnarray}
   \mbox{Coherent States}\quad &&\ket{\alpha}=D(\alpha)\ket{0}
   \quad \mbox{for} \quad \alpha \in \fukuso   \nonumber \\
   \mbox{Squeezed--like States}\quad &&\ket{\beta}=S(\beta)\ket{0}
   \ \quad \mbox{for} \quad \beta \in \fukuso    \nonumber 
\end{eqnarray}

\subsection{Some Useful Formulas}

We list and prove some useful formulas in the following. 
Now we prepare some parameters $\alpha,\ \epsilon,\ \kappa$ in which 
$\epsilon, \kappa$ are free ones, while $\alpha$ is unknown one in the 
cloning case. 
Let us unify the notations as follows. 
\begin{equation}
\label{eq:notations}
 \alpha : \mbox{(unknown)}\quad \alpha=\zettai{\alpha}\mbox{e}^{i\chi};\quad 
 \epsilon : \mbox{known}\quad 
            \epsilon=\zettai{\epsilon}\mbox{e}^{i\phi};\quad 
 \kappa : \mbox{known}\quad  
          \kappa=\zettai{\kappa}\mbox{e}^{i\delta}.
\end{equation}

\par \noindent
\mbox{(i)}\quad  First let us calculate
\begin{equation}
S(\epsilon)D(\alpha)S(\epsilon)^{-1}.
\end{equation}
For that we can show 
\begin{equation}
\label{eq:squeezed-rotation}
S(\epsilon)aS(\epsilon)^{-1}=cosh(\zettai{\epsilon})a
        -\mbox{e}^{i\phi}sinh(\zettai{\epsilon})a^{\dagger}.
\end{equation}

\par \noindent
From this it is easy to check 
\begin{eqnarray}
\label{eq:adjoint-alpha}
 S(\epsilon)D(\alpha)S(\epsilon)^{-1}&=&
 D\left(\alpha S(\epsilon)a^{\dagger}S(\epsilon)^{-1}-
 {\bar \alpha}S(\epsilon)aS(\epsilon)^{-1}\right)
   \nonumber \\
 &=&D\left(cosh(\zettai{\epsilon})\alpha
    +\mbox{e}^{i\phi}sinh(\zettai{\epsilon}){\bar \alpha}\right).  
\end{eqnarray}
Therefore 
\begin{equation}
\label{eq:adjoint-alpha-choice}
S(\epsilon)D(\alpha)S(\epsilon)^{-1}=\left\{
   \begin{array}{ll}
     \displaystyle{D(\mbox{e}^{\zettai{\epsilon}}\alpha)}\ 
       \qquad  \mbox{if}\quad \phi=2\chi   \\
     \displaystyle{D(\mbox{e}^{-\zettai{\epsilon}}\alpha)}
       \qquad  \mbox{if}\quad \phi=2\chi+\pi
   \end{array}
 \right.
\end{equation}

\par \noindent 
This formula is a bit delicate in the cloning case. 
That is, if we could know $\chi$ the phase of $\alpha$ in advance, 
then we can change a scale of $\alpha$ by making use of this one.

\par \vspace{3mm} \noindent
\mbox{(ii)}\quad  Next le us calculate 
\begin{equation}
S(\epsilon)S(\alpha)S(\epsilon)^{-1}.
\end{equation}
From the definition 
\[
 S(\epsilon)S(\alpha)S(\epsilon)^{-1}
 =S(\epsilon)\mbox{exp}\left\{ \frac{1}{2}\left(\alpha (a^{\dagger})^{2}-
 {\bar \alpha}a^{2}\right) \right\}S(\epsilon)^{-1}
 \equiv \mbox{e}^{Y/2}
\]
where 
\[
Y=\alpha\left(S(\epsilon)a^{\dagger}S(\epsilon)^{-1}\right)^{2}
  -{\bar \alpha}\left(S(\epsilon)aS(\epsilon)^{-1}\right)^{2}.
\]
From (\ref{eq:squeezed-rotation}) and after some calculations we have 
\begin{eqnarray}
 Y&=&\left\{ cosh^{2}(\zettai{\epsilon})\alpha
         -\mbox{e}^{2i\phi}sinh^{2}(\zettai{\epsilon}){\bar \alpha}
  \right\}(a^{\dagger})^{2}-
  \left\{ cosh^{2}(\zettai{\epsilon}){\bar \alpha}
         -\mbox{e}^{-2i\phi}sinh^{2}(\zettai{\epsilon})\alpha
  \right\}a^{2}    \nonumber \\
  &+&(-\mbox{e}^{-i\phi}\alpha+\mbox{e}^{i\phi}{\bar \alpha})
    sinh(2\zettai{\epsilon})(a^{\dagger}a+1/2), 
\end{eqnarray}
see (\ref{eq:2-2-12}). This is our second formula. 
\quad Now
\[
  -\mbox{e}^{-i\phi}\alpha+\mbox{e}^{i\phi}{\bar \alpha}
  =\zettai{\alpha}(-\mbox{e}^{-i(\phi-\chi)}+\mbox{e}^{i(\phi-\chi)})
  =2i\zettai{\alpha}sin(\phi-\chi),
\]
so if we choose $\phi=\chi$,\ then $\mbox{e}^{2i\phi}{\bar \alpha}=
\mbox{e}^{2i\chi}\mbox{e}^{-i\chi}\zettai{\alpha}=
\alpha$ and 
\[
   cosh^{2}(\zettai{\epsilon}){\alpha}
         -\mbox{e}^{2i\phi}sinh^{2}(\zettai{\epsilon}){\bar \alpha} 
  =\left(cosh^{2}(\zettai{\epsilon})-
         sinh^{2}(\zettai{\epsilon})\right)\alpha 
  =\alpha,
\]
and finally 
\ \ 
$Y=\alpha(a^{\dagger})^2-{\bar \alpha}a^{2}$.
\ \ 
That is, 
\[
  S(\epsilon)S(\alpha)S(\epsilon)^{-1}=S(\alpha) \Longleftrightarrow 
  S(\epsilon)S(\alpha)=S(\alpha)S(\epsilon).
\]
The operators $S(\epsilon)$ and $S(\alpha)$ commute if the phases of 
$\epsilon$ and $\alpha$ coincide.

\par \vspace{3mm} \noindent
\mbox{(iii)}\ Third formula is : \quad For $V(t)=\mbox{e}^{itN}$\ where 
$N=a^{\dagger}a$ (a number operator) 
\begin{equation}
\label{eq:phase-factor}
V(t)D(\alpha)V(t)^{-1}=D(\mbox{e}^{it}\alpha).
\end{equation}
{\noindent}The proof is as follows. \quad 
\begin{eqnarray}
V(t)D(\alpha)V(t)^{-1}
&=&\mbox{exp}\left(\alpha V(t)a^{\dagger}V(t)^{-1}-
   {\bar \alpha}V(t)aV(t)^{-1}\right)  \nonumber \\
&=&\mbox{exp}\left(\alpha \mbox{e}^{it}a^{\dagger}
   -{\bar \alpha}\mbox{e}^{-it}a\right)
   =D(\mbox{e}^{it}\alpha),
\end{eqnarray}
where we have used 
\[
V(t)aV(t)^{-1}=\mbox{e}^{itN}a\mbox{e}^{-itN}=\mbox{e}^{-it}a.
\]
This formula is often used as follows. 
\begin{equation}
\label{eq:phase-operator}
\ket{\alpha}\ \longrightarrow \ 
V(t)\ket{\alpha}=V(t)D(\alpha)V(t)^{-1}V(t)\ket{0}
=D(\mbox{e}^{it}\alpha)\ket{0}
=\ket{\mbox{e}^{it}\alpha},
\end{equation}
where we have used \ $V(t)\ket{0}=\ket{0}$. 
That is, we can add a phase to $\alpha$ by making use of this formula.

\par \vspace{3mm} \noindent
\mbox{(iv)}\ Fourth formula is : \quad  Let us calculate the following 
\begin{equation}
\label{eq:squeezed-adjoint-formula}
 U_{J}(t)S_{1}(\alpha)S_{2}(\beta)U_{J}(t)^{-1} 
=U_{J}(t)\mbox{e}^{\left\{
   \frac{\alpha}{2}(a_{1}^{\dagger})^{2}-
   \frac{{\bar \alpha}}{2}(a_{1})^{2}
 + \frac{\beta}{2}(a_{2}^{\dagger})^{2}-
   \frac{{\bar \beta}}{2}(a_{2})^{2}
                   \right\}
                  }U_{J}(t)^{-1}=\mbox{e}^{\mbox{X}}
\end{equation}  
where
\begin{eqnarray}
 \mbox{X}&=&
           \frac{\alpha}{2}(U_{J}(t)a_{1}^{\dagger}U_{J}(t)^{-1})^{2}- 
           \frac{{\bar \alpha}}{2}(U_{J}(t)a_{1}U_{J}(t)^{-1})^{2}
     \nonumber \\
 &+&       \frac{\beta}{2}(U_{J}(t)a_{2}^{\dagger}U_{J}(t)^{-1})^{2}- 
           \frac{{\bar \beta}}{2}(U_{J}(t)a_{2}U_{J}(t)^{-1})^{2}. 
     \nonumber 
\end{eqnarray}
From (\ref{eq:J-rotation-1}) and (\ref{eq:J-rotation-2}) we have 
\begin{eqnarray}
\mbox{X}&=&
\frac{1}{2}\left\{cos^{2}(\zettai{t})\alpha + 
\frac{t^2 sin^{2}(\zettai{t})}{\zettai{t}^2}\beta\right\}
    (a_{1}^{\dagger})^2
-\frac{1}{2}\left\{cos^{2}(\zettai{t}){\bar \alpha} + 
\frac{{\bar t}^2 sin^{2}(\zettai{t})}{\zettai{t}^2}{\bar \beta}\right\}
a_{1}^2  \nonumber \\
&+&\frac{1}{2}\left\{cos^{2}(\zettai{t})\beta + 
\frac{{\bar t}^2 sin^{2}(\zettai{t})}{\zettai{t}^2}\alpha\right\}
    (a_{2}^{\dagger})^2
-\frac{1}{2}\left\{cos^{2}(\zettai{t}){\bar \beta} + 
\frac{t^2 sin^{2}(\zettai{t})}{\zettai{t}^2}{\bar \alpha}\right\}a_{2}^2
\nonumber \\
&+&(\beta t-\alpha {\bar t})\frac{sin(2\zettai{t})}{2\zettai{t}}
      a_{1}^{\dagger}a_{2}^{\dagger}
-({\bar \beta}{\bar t}-{\bar \alpha}t)\frac{sin(2\zettai{t})}{2\zettai{t}}
      a_{1}a_{2}. 
\end{eqnarray}
If we set 
\begin{equation}
\label{eq:strong-condition}
\beta t-\alpha {\bar t}=0 \Longleftrightarrow \beta t=\alpha {\bar t}, 
\end{equation}
then it is easy to check 
\[
cos^{2}(\zettai{t})\alpha + 
\frac{t^2 sin^{2}(\zettai{t})}{\zettai{t}^2}\beta=\alpha, 
\quad 
cos^{2}(\zettai{t})\beta + 
\frac{{\bar t}^2 sin^{2}(\zettai{t})}{\zettai{t}^2}\alpha=\beta, 
\]
so, in this case, 
\[
X=\frac{1}{2}\alpha(a_{1}^{\dagger})^2-\frac{1}{2}{\bar \alpha}a_{1}^2 + 
  \frac{1}{2}\beta(a_{2}^{\dagger})^2-\frac{1}{2}{\bar \beta}a_{2}^2\ . 
\]
Therefore 
\begin{equation}
\label{eq:2-invariant-property}
U_{J}(t)S_{1}(\alpha)S_{2}(\beta)U_{J}(t)^{-1}=
S_{1}(\alpha)S_{2}(\beta).
\end{equation}
That is, $S_{1}(\alpha)S_{2}(\beta)$ commutes with $U_{J}(t)$ under the 
condition (\ref{eq:strong-condition}).

\subsection{Swap of Coherent States}
The purpose of this section is to construct a swap operator 
satifying 
\begin{equation}
     \ket{\alpha_{1}}\otimes \ket{\alpha_{2}} \longrightarrow 
     \ket{\alpha_{2}}\otimes \ket{\alpha_{1}}.
\end{equation}
Let us remember $U_{J}(\kappa)$ once more 
\[
U_{J}(\kappa)=\mbox{e}^{\kappa a_{1}^{\dagger}a_{2}-{\bar \kappa}
a_{1}a_{2}^{\dagger}} \quad \mbox{for} \quad \kappa \in \fukuso .
\]
We note an important property of this operator : 
\begin{equation}
\label{eq:invariant-property}
U_{J}(\kappa)\ket{0}\otimes \ket{0}=\ket{0}\otimes \ket{0}.
\end{equation}
The construction is as follows. 
\begin{eqnarray}
\label{eq:swap-operator}
U_{J}(\kappa)\ket{\alpha_{1}}\otimes \ket{\alpha_{2}}
&=&U_{J}(\kappa)D(\alpha_{1})\otimes D(\alpha_{2})\ket{0}\otimes \ket{0}
=U_{J}(\kappa)D_{1}(\alpha_{1})D_{2}(\alpha_{2})\ket{0}\otimes \ket{0}
      \nonumber \\
&=&U_{J}(\kappa)D_{1}(\alpha_{1})D_{2}(\alpha_{2})U_{J}(\kappa)^{-1}
\ket{0}\otimes \ket{0} \quad \mbox{by}\quad (\ref{eq:invariant-property}), 
\end{eqnarray}
and 
\begin{eqnarray}
&&U_{J}(\kappa)D_{1}(\alpha_{1})D_{2}(\alpha_{2})U_{J}(\kappa)^{-1}=
U_{J}(\kappa)
\mbox{exp}\left\{\alpha_{1}a_{1}^{\dagger}-{\bar \alpha_{1}}a_{1} + 
          \alpha_{2}a_{2}^{\dagger}-{\bar \alpha_{2}}a_{2}
          \right\}
U_{J}(\kappa)^{-1}  \nonumber \\
&&=\mbox{exp}
\left\{
\alpha_{1}(U_{J}(\kappa)a_{1}U_{J}(\kappa)^{-1})^{\dagger}-
{\bar \alpha_{1}}U_{J}(\kappa)a_{1}U_{J}(\kappa)^{-1} 
\right.  \nonumber \\
&&\left. \qquad \ + 
\alpha_{2}(U_{J}(\kappa)a_{2}U_{J}(\kappa)^{-1})^{\dagger}-
{\bar \alpha_{2}}U_{J}(\kappa)a_{2}U_{J}(\kappa)^{-1}
\right\}  \nonumber \\
     \label{eq:tensor-product}
&&\equiv \mbox{exp}(X).
\end{eqnarray}
From (\ref{eq:J-rotation-1}) and (\ref{eq:J-rotation-2}) we have 
\begin{eqnarray}
X
&=&
\left\{
cos(\zettai{\kappa})\alpha_{1}+
\frac{\kappa sin(\zettai{\kappa})}{\zettai{\kappa}}\alpha_{2}
\right\}a_{1}^{\dagger}
-
\left\{
cos(\zettai{\kappa}){\bar \alpha_{1}}+
\frac{{\bar \kappa}sin(\zettai{\kappa})}{\zettai{\kappa}}{\bar \alpha_{2}}
\right\}a_{1}  \nonumber \\
&+&
\left\{
cos(\zettai{\kappa})\alpha_{2}-
\frac{{\bar \kappa}sin(\zettai{\kappa})}{\zettai{\kappa}}\alpha_{1}
\right\}a_{2}^{\dagger}
-
\left\{
cos(\zettai{\kappa}){\bar \alpha_{2}}-
\frac{\kappa sin(\zettai{\kappa})}{\zettai{\kappa}}{\bar \alpha_{1}}
\right\}a_{2},  \nonumber
\end{eqnarray}
so
\begin{eqnarray}
\mbox{exp}(X)
&=&
D_{1}
\left(
cos(\zettai{\kappa})\alpha_{1}+
\frac{\kappa sin(\zettai{\kappa})}{\zettai{\kappa}}\alpha_{2}
\right) 
D_{2}
\left(
cos(\zettai{\kappa})\alpha_{2}-
\frac{{\bar \kappa}sin(\zettai{\kappa})}{\zettai{\kappa}}\alpha_{1}
\right)  \nonumber \\
&=&
D
\left(
cos(\zettai{\kappa})\alpha_{1}+
\frac{\kappa sin(\zettai{\kappa})}{\zettai{\kappa}}\alpha_{2}
\right)
\otimes
D
\left(
cos(\zettai{\kappa})\alpha_{2}-
\frac{{\bar \kappa}sin(\zettai{\kappa})}{\zettai{\kappa}}\alpha_{1}
\right).  \nonumber 
\end{eqnarray}
Therefore we have from (\ref{eq:tensor-product})
\[
\ket{\alpha_{1}}\otimes \ket{\alpha_{2}}\  \longrightarrow \ 
\ket{
cos(\zettai{\kappa})\alpha_{1}+
\frac{\kappa sin(\zettai{\kappa})}{\zettai{\kappa}}\alpha_{2}
} 
\otimes 
\ket{
cos(\zettai{\kappa})\alpha_{2}-
\frac{{\bar \kappa}sin(\zettai{\kappa})}{\zettai{\kappa}}\alpha_{1}
}.
\]
If we write $\kappa$ as $\zettai{\kappa}\mbox{e}^{i\delta}$ from 
(\ref{eq:notations}), then the above formula reduces to
\begin{equation}
\label{eq:formula-formula}
\ket{\alpha_{1}}\otimes \ket{\alpha_{2}}\  \longrightarrow \ 
\ket{
cos(\zettai{\kappa})\alpha_{1}+
\mbox{e}^{i\delta}sin(\zettai{\kappa})\alpha_{2}
}
\otimes
\ket{
cos(\zettai{\kappa})\alpha_{2}-
\mbox{e}^{-i\delta}sin(\zettai{\kappa})\alpha_{1}
}.
\end{equation}
This is a central formula. 
Here if we choose $sin(\zettai{\kappa})=1$, then 
\[
\ket{\alpha_{1}}\otimes \ket{\alpha_{2}}\  \longrightarrow \ 
\ket{\mbox{e}^{i\delta}\alpha_{2}}
\otimes
\ket{-\mbox{e}^{-i\delta}\alpha_{1}}
=
\ket{\mbox{e}^{i\delta}\alpha_{2}}
\otimes
\ket{\mbox{e}^{-i(\delta+\pi)}\alpha_{1}}.
\]
Now by operating the operator $V=\mbox{e}^{-i\delta N}\otimes 
\mbox{e}^{i(\delta+\pi) N}$ where $N=a^{\dagger}a$ 
from the left (see (\ref{eq:phase-operator})) 
we obtain the swap 
\[
\ket{\alpha_{1}}\otimes \ket{\alpha_{2}}\  \longrightarrow \ 
\ket{\alpha_{2}}\otimes \ket{\alpha_{1}}.
\]

\par \noindent
{\bf A comment is in order}.\ \ In the formula (\ref{eq:formula-formula}) 
we set $\alpha_{1}=\alpha$ and $\alpha_{2}=0$, 
then (\ref{eq:formula-formula}) reduces to 
\begin{equation}
\label{eq:adjoint-form}
\ket{\alpha}\otimes \ket{0} \longrightarrow 
 \ket{cos(\zettai{\kappa})\alpha}\otimes 
 \ket{-\mbox{e}^{-i\delta}sin(\zettai{\kappa})\alpha}
=\ket{cos(\zettai{\kappa})\alpha}\otimes 
 \ket{\mbox{e}^{-i(\delta+\pi)}sin(\zettai{\kappa})\alpha}.
\end{equation}

\subsection{Imperfect Cloning of Coherent States}
We cannot clone coherent states in a perfect manner like 
\begin{equation}
   \ket{\alpha}\otimes \ket{0} \longrightarrow 
   \ket{\alpha}\otimes \ket{\alpha} \quad \mbox{for}\quad 
   \alpha \in \fukuso .
\end{equation}
Then our question is : is it possible to approximate ?  
Here let us note once more that $\alpha$ is in this case unknown. 
We show that we can at least make an ``imperfect cloning" in our terminology 
against the statement of \cite{DG}. 
The method is almost same with one in the preceding subsection. 
By (\ref{eq:adjoint-form}) 
\[
\ket{\alpha}\otimes \ket{0} \longrightarrow 
 \ket{cos(\zettai{\kappa})\alpha}\otimes 
 \ket{\mbox{e}^{-i(\delta+\pi)}sin(\zettai{\kappa})\alpha}.
\]
we have by operating the operator ${\bf 1}\otimes \mbox{e}^{i(\delta+\pi)N}$ 
(see (\ref{eq:phase-operator})) 
\begin{equation}
  \ket{\alpha}\otimes \ket{0} \longrightarrow 
 \ket{cos(\zettai{\kappa})\alpha}\otimes 
 \ket{sin(\zettai{\kappa})\alpha}. 
\end{equation}
Here if we set $\zettai{\kappa}=\pi/4$ in particular, then we have 
\begin{equation}
  \ket{\alpha}\otimes \ket{0} \longrightarrow 
 \ket{\frac{\alpha}{\sqrt{2}}}\otimes 
 \ket{\frac{\alpha}{\sqrt{2}}}. 
\end{equation}
This is the ``imperfect cloning" which we have called. 

\par \noindent
{\bf A comment is in order.} \quad 
The authors in \cite{DG} state that the ``perfect cloning" (in their 
terminology) for coherent states is possible. But it is not correct as 
shown in \cite{KF7}. 
Nevertheless their method is simple and very interesting, so it 
may be possible to modify their ``proof" more subtly by making use of 
(\ref{eq:adjoint-alpha-choice}).

\subsection{Swap of Squeezed--like States ?}
We would like to construct an operator like 
\begin{equation}
     \ket{\beta_{1}}\otimes \ket{\beta_{2}} \longrightarrow 
     \ket{\beta_{2}}\otimes \ket{\beta_{1}}. 
\end{equation}
In this case we cannot use an operator $U_{J}(\kappa)$. 
Let us explain the reason. 
Similar to (\ref{eq:swap-operator}) 
\begin{eqnarray}
U_{J}(\kappa)\ket{\beta_{1}}\otimes \ket{\beta_{2}}
&=&U_{J}(\kappa)S(\beta_{1})\otimes S(\beta_{2})\ket{0}\otimes \ket{0}
 =U_{J}(\kappa)S_{1}(\beta_{1})S_{2}(\beta_{2})\ket{0}\otimes \ket{0}
       \nonumber \\
&=&U_{J}(\kappa)S_{1}(\beta_{1})S_{2}(\beta_{2})U_{J}(\kappa)^{-1}
\ket{0}\otimes \ket{0}.
\end{eqnarray}
On the other hand by (\ref{eq:squeezed-adjoint-formula}) 
\[
U_{J}(\kappa)S_{1}(\beta_{1})S_{2}(\beta_{2})U_{J}(\kappa)^{-1}
=\mbox{e}^{X},
\]
where 
\begin{eqnarray}
\mbox{X}&=&
\frac{1}{2}\left\{cos^{2}(\zettai{\kappa})\beta_{1} + 
\frac{\kappa^2 sin^{2}(\zettai{\kappa})}{\zettai{\kappa}^2}\beta_{2}\right\}
    (a_{1}^{\dagger})^2
-\frac{1}{2}\left\{cos^{2}(\zettai{\kappa}){\bar \beta_{1}} + 
\frac{{\bar \kappa}^2 sin^{2}(\zettai{\kappa})}{\zettai{\kappa}^2}
{\bar \beta_{2}}\right\}a_{1}^2       \nonumber \\
&+&\frac{1}{2}\left\{cos^{2}(\zettai{\kappa})\beta_{2} + 
\frac{{\bar \kappa}^2 sin^{2}(\zettai{\kappa})}{\zettai{\kappa}^2}
\beta_{1}\right\}(a_{2}^{\dagger})^2
-\frac{1}{2}\left\{cos^{2}(\zettai{\kappa}){\bar \beta_{2}} + 
\frac{\kappa^2 sin^{2}(\zettai{\kappa})}{\zettai{\kappa}^2}
{\bar \beta_{1}}\right\}a_{2}^2        \nonumber \\
&+&(\beta_{2} \kappa-\beta_{1} {\bar \kappa})
\frac{sin(2\zettai{\kappa})}{2\zettai{\kappa}}a_{1}^{\dagger}a_{2}^{\dagger}
-({\bar \beta_{2}}{\bar \kappa}-{\bar \beta_{1}}\kappa)
\frac{sin(2\zettai{\kappa})}{2\zettai{\kappa}}a_{1}a_{2}\ .   \nonumber 
\end{eqnarray}
Here an extra term containing $a_{1}^{\dagger}a_{2}^{\dagger}$ appeared. 
To remove this we must set $\beta_{2} \kappa-\beta_{1} {\bar \kappa}=0$, 
but in this case we meet 
\[
U_{J}(\kappa)S_{1}(\beta_{1})S_{2}(\beta_{2})U_{J}(\kappa)^{-1}
=S_{1}(\beta_{1})S_{2}(\beta_{2})
\]
by (\ref{eq:2-invariant-property}). That is, there is no change. 

We could not construct operators as in the subsection 3.2 in spite of 
very our efforts, so we present 

\par \noindent
\begin{flushleft}
{\bf Problem}\quad 
Is it possible to find an operator such as $U_{J}(\kappa)$ in the preceding 
subsection for performing the swap ? 
\end{flushleft}

\subsection{Squeezed--Coherent States}
We introduce interesting states called squeezed--coherent ones : 
\begin{equation}
     \ket{(\beta,\alpha)}=S(\beta)D(\alpha)\ket{0}\quad \mbox{for}\quad 
     \beta,\ \alpha \in \fukuso.
\end{equation}
$\ket{(\beta,0)}$ is a squeezed--like state and $\ket{(0,\alpha)}$ is 
a coherent one. 
These states play a very important role in Holonomic Quantum Computation, 
see for example \cite{KF00}, \cite{KF01} or \cite{ZR}, \cite{PZR}. 

\par \noindent
\begin{flushleft}
{\bf Problem}\quad 
Is it possible to find some operators for performing the swap or 
imperfect cloning ? 
\end{flushleft}

\par \vspace{5mm} \noindent 

\begin{center}
\begin{Large}
{\bf Appendix \quad Universal Swap Operator}
\end{Large}
\end{center}

\par \vspace{3mm} \noindent
Let us construct the swap operator in a universal manner 
\[
U : \calh\otimes \calh \longrightarrow \calh\otimes \calh \ ,\quad 
    U(a\otimes b)=b\otimes a \quad \mbox{for}\quad a,\ b \in \calh 
\]
where $\calh$ is an infinite--dimensional Hilbert space. Before constructing 
it we show in the finite--dimensional case, \cite{KF8}.

For $a,\ b \in {\fukuso}^2$ then
\[
a\otimes b=
\left(
\begin{array}{c}
a_{1}b \\
a_{2}b
\end{array}
\right)
=
\left(
\begin{array}{c}
a_{1}b_{1} \\
a_{1}b_{2} \\
a_{2}b_{1} \\
a_{2}b_{2} 
\end{array}
\right),
\quad 
b\otimes a=
\left(
\begin{array}{c}
b_{1}a_{1} \\
b_{1}a_{2} \\
b_{2}a_{1} \\
b_{2}a_{2} 
\end{array}
\right)
=
\left(
\begin{array}{c}
a_{1}b_{1} \\
a_{2}b_{1} \\
a_{1}b_{2} \\
a_{2}b_{2} 
\end{array}
\right),
\]
so it is easy to see
\[
\left(
\begin{array}{cccc}
1& 0& 0& 0 \\
0& 0& 1& 0 \\
0& 1& 0& 0 \\
0& 0& 0& 1
\end{array}
\right)
\left(
\begin{array}{c}
a_{1}b_{1} \\
a_{1}b_{2} \\
a_{2}b_{1} \\
a_{2}b_{2} 
\end{array}
\right)
=
\left(
\begin{array}{c}
a_{1}b_{1} \\
a_{2}b_{1} \\
a_{1}b_{2} \\
a_{2}b_{2} 
\end{array}
\right).
\]
That is, the swap operator is 
\begin{equation}
U=
\left(
\begin{array}{cccc}
1& 0& 0& 0 \\
0& 0& 1& 0 \\
0& 1& 0& 0 \\
0& 0& 0& 1
\end{array}
\right). 
\end{equation}
This matrix can be written as follows by making use of three 
Controlled--NOT matrices (gates)
\begin{equation}
\left(
\begin{array}{cccc}
1& 0& 0& 0 \\
0& 0& 1& 0 \\
0& 1& 0& 0 \\
0& 0& 0& 1
\end{array}
\right) 
=
\left(
\begin{array}{cccc}
1& 0& 0& 0 \\
0& 1& 0& 0 \\
0& 0& 0& 1 \\
0& 0& 1& 0
\end{array}
\right) 
\left(
\begin{array}{cccc}
1& 0& 0& 0 \\
0& 0& 0& 1 \\
0& 0& 1& 0 \\
0& 1& 0& 0
\end{array}
\right) 
\left(
\begin{array}{cccc}
1& 0& 0& 0 \\
0& 1& 0& 0 \\
0& 0& 0& 1 \\
0& 0& 1& 0
\end{array}
\right),
\end{equation}
See for example \cite{KF1}. 

\par \noindent
It is not easy for us to conjecture its general form from this 
swap operator. Let us try for $n=3$. The result is 
\[
\left(
\begin{array}{ccccccccc}
1& 0& 0& 0& 0& 0& 0& 0& 0 \\
0& 0& 0& 1& 0& 0& 0& 0& 0 \\
0& 0& 0& 0& 0& 0& 1& 0& 0 \\
0& 1& 0& 0& 0& 0& 0& 0& 0 \\
0& 0& 0& 0& 1& 0& 0& 0& 0 \\
0& 0& 0& 0& 0& 0& 0& 1& 0 \\
0& 0& 1& 0& 0& 0& 0& 0& 0 \\
0& 0& 0& 0& 0& 1& 0& 0& 0 \\
0& 0& 0& 0& 0& 0& 0& 0& 1 
\end{array}
\right)
\left(
\begin{array}{c}
a_{1}b_{1} \\
a_{1}b_{2} \\
a_{1}b_{3} \\
a_{2}b_{1} \\
a_{2}b_{2} \\
a_{2}b_{3} \\
a_{3}b_{1} \\
a_{3}b_{2} \\
a_{3}b_{3} 
\end{array}
\right)
=
\left(
\begin{array}{c}
a_{1}b_{1} \\
a_{2}b_{1} \\
a_{3}b_{1} \\
a_{1}b_{2} \\
a_{2}b_{2} \\
a_{3}b_{2} \\ 
a_{1}b_{3} \\
a_{2}b_{3} \\
a_{3}b_{3} 
\end{array}
\right).
\]
Here we rewrite the swap operator above as follows.
\begin{equation}
U=
\left(
\begin{array}{ccc}
 \left(
  \begin{array}{ccc}
   1& 0& 0\\
   0& 0& 0\\
   0& 0& 0
  \end{array}
 \right) &
 \left(
  \begin{array}{ccc}
   0& 0& 0\\
   1& 0& 0\\
   0& 0& 0
  \end{array}
 \right) &
 \left(
  \begin{array}{ccc}
   0& 0& 0\\
   0& 0& 0\\
   1& 0& 0
  \end{array}
 \right) \\
  \left(
  \begin{array}{ccc}
   0& 1& 0\\
   0& 0& 0\\
   0& 0& 0
  \end{array}
 \right) &
 \left(
  \begin{array}{ccc}
   0& 0& 0\\
   0& 1& 0\\
   0& 0& 0
  \end{array}
 \right) &
 \left(
  \begin{array}{ccc}
   0& 0& 0\\
   0& 0& 0\\
   0& 1& 0
  \end{array}
 \right) \\
 \left(
  \begin{array}{ccc}
   0& 0& 1\\
   0& 0& 0\\
   0& 0& 0
  \end{array}
 \right) &
 \left(
  \begin{array}{ccc}
   0& 0& 0\\
   0& 0& 1\\
   0& 0& 0
  \end{array}
 \right) &
 \left(
  \begin{array}{ccc}
   0& 0& 0\\
   0& 0& 0\\
   0& 0& 1
  \end{array}
 \right) 
\end{array}
\right).
\end{equation}
Now, from the above form 
we can conjecture the general form of the swap operator.

We note that 
\begin{equation}
({\bf 1}\otimes {\bf 1})_{ij,kl}=\delta_{ik}\delta_{jl},
\end{equation}
so after some trials we conclude 
\[
U : {\fukuso}^{n}\otimes {\fukuso}^{n} \longrightarrow 
    {\fukuso}^{n}\otimes {\fukuso}^{n}
\]
as 
\begin{equation}
U=(U_{ij,kl})\quad ;\quad U_{ij,kl}=\delta_{il}\delta_{jk},
\end{equation}
where $ij=11,12,\cdots,1n, 21,22,\cdots,2n,\ \cdots,\ n1,n2,\cdots,nn$.

The proof is simple and as follows.\quad 
\begin{eqnarray}
(a\otimes b)_{ij}=a_{i}b_{j} \longrightarrow 
\left\{U(a\otimes b)\right\}_{ij}
&=&\sum_{kl=11}^{nn}U_{ij,kl}a_{k}b_{l}
=\sum_{kl=11}^{nn}\delta_{il}\delta_{jk}a_{k}b_{l} \nonumber \\
&=&\sum_{l=1}^{n}\delta_{il}b_{l} \sum_{k=1}^{n}\delta_{jk}a_{k}
=b_{i}a_{j}=(b\otimes a)_{ij}.  \nonumber 
\end{eqnarray}
At this stage there is no problem to take a limit $n \rightarrow \infty$. 

Let $\calh$ be a Hilbert space with a basis $\{e_{n}\}$ ($n \geq 1$). 
Then the universal swap operator is given by
\begin{equation}
U=(U_{ij,kl})\quad ;\quad U_{ij,kl}=\delta_{il}\delta_{jk},
\end{equation}
where $ij=11,12,\cdots, \cdots$.

\par \noindent
We note that this is not a physical construction but only 
a mathematical (abstract) one, so we have a natural question.

\par \vspace{3mm} \noindent
{\bf Problem}\quad 
Is it possible to realize this universal swap operator in Quantum Optics ?


\end{document}